\documentclass{emulateapj}
\newcommand{\etal}{et al.}

\newcommand\xte{{\it RXTE\/}}
\newcommand\chandra{{\it Chandra}}
\newcommand\xmm{{\it XMM-Newton}}

\def\psr{\hbox{XTE J1829$-$098}}
\def\lpv{\hbox{2MASS J182944.55$-$095120.3}}

\slugcomment{To appear in The Astrophysical Journal}

\shorttitle{X-ray and IR Observations of XTE J1829$-$098}
\shortauthors{Halpern and Gotthelf}

\begin{document}

\title{X-ray Observations and Infrared Identification of the Transient\\
7.8~s X-ray Binary Pulsar XTE J1829$-$098}

\author{J. P. Halpern and E. V. Gotthelf}

\affil{Columbia Astrophysics Laboratory,
Columbia University, 550 West 120th Street, New York, NY 10027}

\begin{abstract}
\xmm\ and \chandra\ observations of the transient 7.8~s pulsar 
\psr\ are used to characterize its pulse shape and spectrum, and
to facilitate a search for an optical or infrared counterpart.
In outburst, the
absorbed, hard X-ray spectrum with $\Gamma = 0.76\pm 0.13$
and $N_{\rm H} = (6.0 \pm 0.6) \times 10^{22}$~cm$^{-2}$
is typical of X-ray binary pulsars.  The precise \chandra\
localization in a faint state leads to the identification of
a probable infrared
counterpart at R.A. = $18^{\rm h}29^{\rm m}43.\!^{\rm s}98$,
decl. = $-09^{\circ}51^{\prime}23^{\prime\prime}\!.0$ (J2000.0)
with magnitudes $K=12.7$, $H=13.9$, $I>21.9$, and $R>23.2$.
If this is a highly reddened O or B star, we estimate
a distance of 10~kpc, at which the maximum observed X-ray luminosity
is $2 \times 10^{36}$ ergs~s$^{-1}$, typical of Be X-ray transients
or wind-fed systems.  The minimum observed luminosity is
$3 \times 10^{32}(d/10\ {\rm kpc})^2$ ergs~s$^{-1}$.
We cannot rule out the possibility
that the companion is a red giant.
The two known X-ray outbursts of \psr\
are separated by $\approx 1.3$ yr, which may be the orbital period
or a multiple of it, with the neutron star in an eccentric orbit.
We also studied a late M-giant long-period variable
that we found only $9^{\prime\prime}$ from the X-ray position.  It
has a pulsation period of $\approx 1.5$~yr, but is
not the companion of the X-ray source.

\end{abstract}
\keywords{pulsars: individual (\psr) --- stars: neutron --- 
stars: AGB and post-AGB --- stars: individual (\lpv) ---
stars: variable: other --- X-rays: binaries}

\section {Introduction}

Scans of the Galactic plane in 2004 July-August with the
{\it Rossi X-ray Timing Explorer} \citep[\xte;][]{bra93}
Proportional Counter Array \citep[PCA;][]{jah97}
detected a new, transient 7.8~s X-ray pulsar (Markwardt et al. 2004).
Those authors noted that the source's hard X-ray
spectrum favors classification as a high-mass X-ray binary (HMXB)
rather than a transient Anomalous X-ray Pulsar (AXP), although a
low-mass X-ray binary (LMXB) pulsar similar to 4U~1626$-$67
could not be ruled out.
We determined that three pointings had been made previously at this
location by the {\it Newton X-ray Multi-Mirror Mission} (\xmm),
and that a bright source was present
in one of these unpublished observations from 2003 March.
A preliminary analysis
clearly revealed its identity with \psr\ by
virtue of its 7.8~s pulsations and X-ray spectral parameters
compatible with the \xte\ measured ones (Halpern \& Gotthelf 2004a).
The repeated outburst of the source, separated by non-detections in
\xmm\ observations at a level at least 3400 times fainter,
is typical behavior of a transient HMXB.

In \S 2, we describe the \xmm\ observation of \psr\ in its
outburst state, as well as upper limits from non-detections.
In \S 3, we report \chandra\ observations that localize the
source in its faint state, followed by optical and infrared
imaging in \S 4 that identifies a reddened counterpart.
The still uncertain classification of \psr\ is discussed in \S 5.
In the Appendix, we present photometry and spectroscopy of
a bright, newly recognized long-period variable star that is only
$9^{\prime\prime}$ from the X-ray source.

\begin{deluxetable*}{llcccc}[t]
\tablecaption{X-ray Observations of \psr}
\tablehead
{\colhead{Mission} & \colhead{Date} &
\colhead{$F_x(2-10\ {\rm keV})$\tablenotemark{a}} & \colhead{$\Gamma$} &
\colhead{$N_{\rm H}$} & \colhead{References} \\
& & \colhead{(ergs cm$^{-2}$ s$^{-1}$)} & &
\colhead{($10^{22}$ cm$^{-2}$)} &
}
\startdata
{\it XMM} & 2002 Mar 27 & $< 5   \times 10^{-14}$ &...&...& 1 \\
{\it XMM} & 2003 Mar 27 & $  8.2 \times 10^{-11}$
& $0.76 \pm 0.13$\tablenotemark{b} 
& $6.0 \pm 0.6$\tablenotemark{b} & 1 \\
{\it XMM} & 2003 Sep 13  & $< 5   \times 10^{-14}$ &...&...& 1 \\
\xte & 2004 May 22 -- Jun 10 & $< 4.8   \times 10^{-11}$ &...&...& 2 \\
\xte & 2004 Jul 30 & $1.7   \times 10^{-10}$ &...&...& 2 \\   
\xte & 2004 Aug 8  & $1.0   \times 10^{-10}$ & 1.0 & 10 & 2 \\
\chandra & 2007 Feb 14 & $2.5 \times 10^{-14}$ &...&...& 1 \\
\chandra & 2007 May 24 & $2.8 \times 10^{-13}$ &...&...& 1
\enddata
\tablenotetext{a}{Absorbed flux, assuming 1~mCrab $= 2.4 \times 10^{-11}$
ergs~cm$^{-2}$~s$^{-1}$ to convert from \citet{mar04}.}
\tablenotetext{b}{Uncertainties are 68\% confidence
for two interesting parameters (see Fig.~\ref{xrayspectrum}).}
\tablerefs{(1) this work;
(2) Markwardt et al. (2004).}
\label{xraytable}
\end{deluxetable*}

\section{XMM-Newton Observations}

\begin{figure}[b]
\centerline{
\includegraphics[width=3.2in,angle=-90.0]{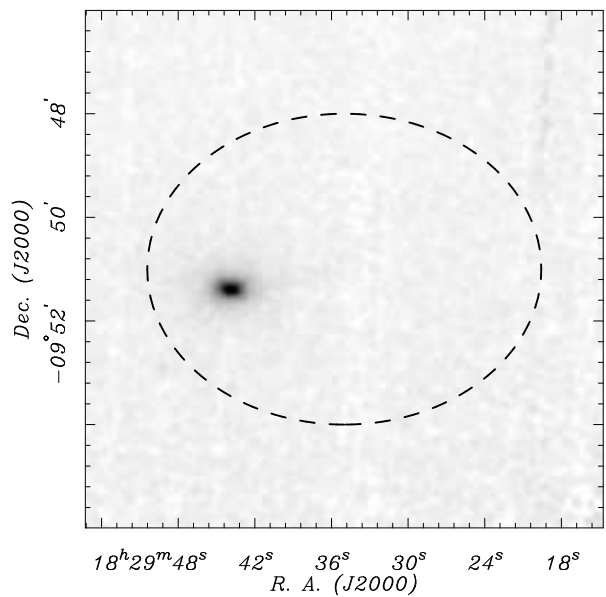}
}
\caption{\small Portion of the \xmm\ image of 2003 March 2007
centered on the \xte\ error ellipse of \psr\ \citep{mar04}.
The EPIC-pn and two EPIC MOS images have been combined.
}
\label{xmmimage}
\end{figure}

Three \xmm\ pointings that include the location of
\psr\ were obtained in 2002 and 2003 as part of the Galactic Plane Survey
\citep{han04}.  We reprocessed them using the emchain and epchain
scripts under Science Analysis System (SAS) version
xmmsas\_20060628\_1801-7.0.0.
Table~\ref{xraytable} lists the basic results
of these observations.
An X-ray source corresponding to \psr\ was detected only during
the 2.6~ks observation of 2003 March 27 (see Fig.~\ref{xmmimage}),
while the 2000 March 27 and 2003 September 13 images, having
exposure times of 6.8~ks and 6.4~ks, respectively, are blank at this
position.  Thus, variability by at least a factor of 3400 is
indicated.  The position of the source is
R.A = $18^{\rm h}29^{\rm m}44.\!^{\rm s}10$,
decl. = $-09^{\circ}51^{\prime}24.\!^{\prime\prime}1$ (J2000.0)
with a nominal 90\% uncertainty radius of $3.\!^{\prime\prime}2$.
This falls within the \xte\ error circle, and we confirm the
identification with \psr\ by finding the corresponding periodic signal at 
barycentric period $P = 7.840 \pm 0.004$~s,
consistent with $P = 7.82 \pm 0.05$~s from \xte\ \citep{mar04}.

\begin{figure}[b]
\centerline{
\includegraphics[width=3.2in]{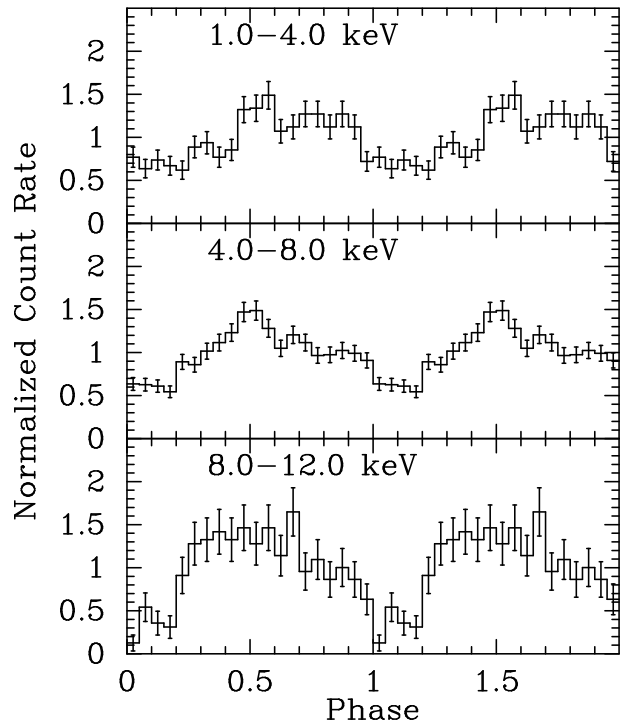}
}
\caption{Folded pulse profiles of \psr\ from the \xmm\
EPIC-pn observation of 2003 March 27.  Background is
subtracted and count rates are
normalized so that the average per bin is 1.
}
\label{xraypulse}
\end{figure}

\begin{figure}[t]
\centerline{
\includegraphics[width=2.2in,angle=-90.0]{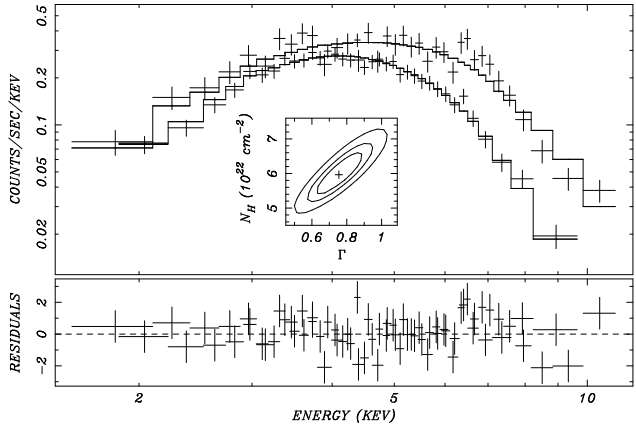}
}
\caption{X-ray spectrum of \psr\ from the \xmm\
observation of 2003 March 27, fitted to
an absorbed power law. The upper curve is from the EPIC-pn,
and the lower curve is from the combined EPIC-MOS detectors.
Residuals are in units of $\sigma$.
{\it Inset\/}: Confidence contours of spectral parameters
are $1\sigma, 2\sigma, 3\sigma$ for two interesting parameters.
}
\label{xrayspectrum}
\end{figure}

The 73.4~ms sampling of the European Photon Imaging Camera
\citep[EPIC-pn;][]{tur03} data amply resolves the pulsations.
Figure~\ref{xraypulse} shows the EPIC-pn pulse profiles at several
energies from 1.0--12~keV.  The pulsed fraction is dependent on
energy, ranging from $\approx 35\%$ at 1~keV to $\approx 70\%$ at 10~keV.
We also fitted the spectra of 2003 March 27, jointly
from the EPIC-pn and the two EPIC MOS CCDs, to an absorbed
power law.  The resulting fit is acceptable ($\chi_{\nu}^2 = 1.0$
for 68 degrees-of-freedom) as shown in
Figure~\ref{xrayspectrum}, yielding a hard power law of photon
index $\Gamma = 0.76 \pm 0.13$ and
$N_{\rm H} = (6.0 \pm 0.6) \times 10^{22}$~cm$^{-2}$
($1\sigma$ uncertainties).
These results are typical of HMXBs.  In this case,
the measured $N_{\rm H}$ exceeds the Galactic 21~cm \ion{H}{1} 
column density of $1.8 \times 10^{22}$~cm$^{-2}$ in this direction
\citep{dic90},
suggesting that some of the absorption is intrinsic to the binary.
However, the molecular hydrogen equivalent of the CO column density
in this direction is $N_H = 2N_{H_2} \sim 5 \times 10^{22}$~cm$^{-2}$
\citep{cle86}
which, together with \ion{H}{1}, could account for the X-ray
measured column density, but only if \psr\ is considerably
farther than the Galactic center.

\begin{figure}[b]
\vspace{0.1in}
\centerline{
\includegraphics[width=1.65in]{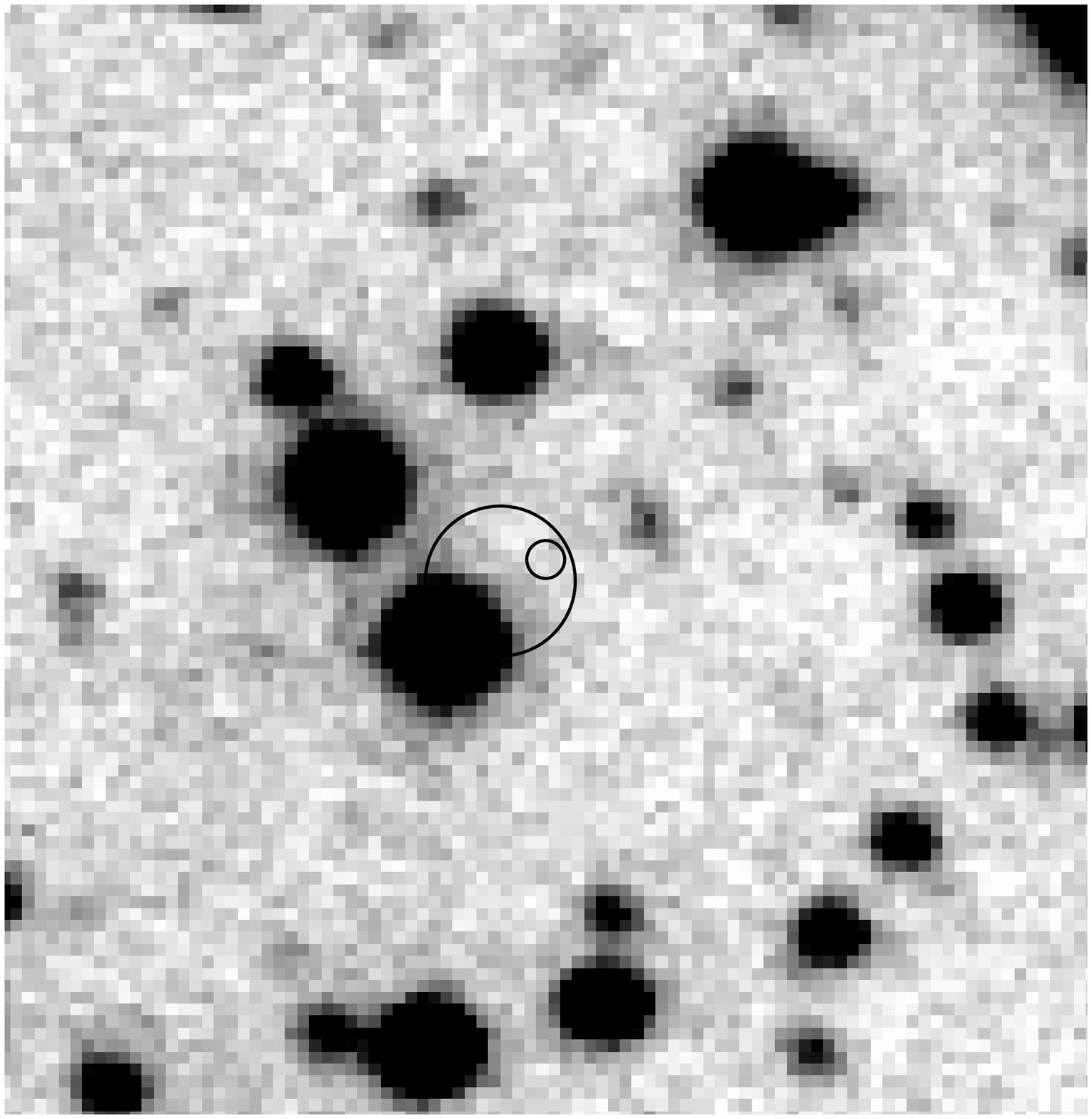}
\includegraphics[width=1.65in]{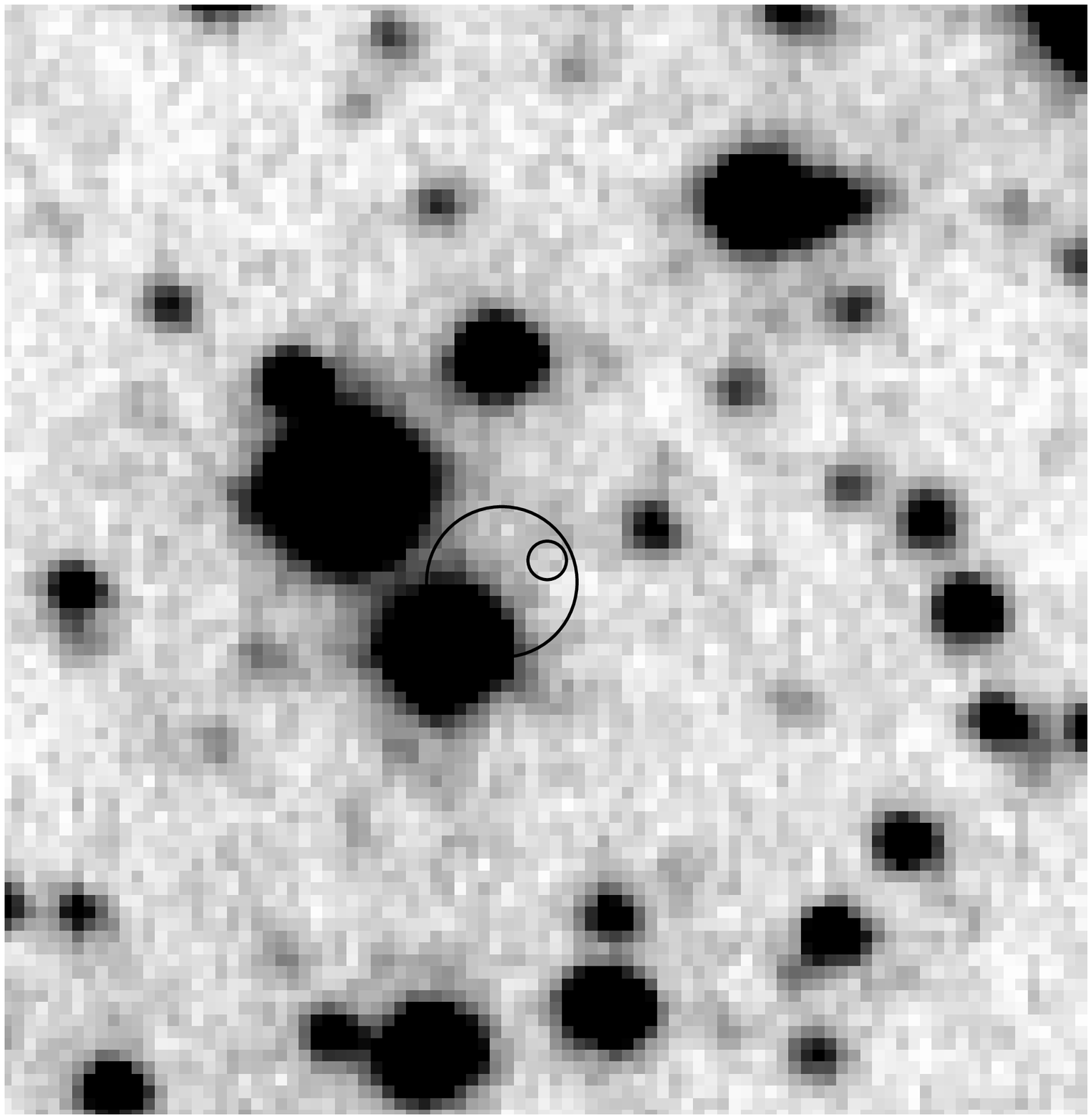}
}
\caption{Combined images of the field of \psr\ from the MDM 1.3m 
telescope.
{\it Left\/}: $R$-band, limiting magnitude 23.2.
{\it Right\/}: $I$-band, limiting magnitude 21.9.
North is up, and east is to the left.
Each image is $0.\!^{\prime}8 \times 0.\!^{\prime}8$.
The large circle is the \xmm\ 90\% localization, of radius
$3.\!^{\prime\prime}2$, and the small circle is the
\chandra\ 90\% localization, of radius $0.\!^{\prime\prime}6$.
The brightest star in the $I$-band image is \lpv,
a late M giant and LPV,
with optical spectrum shown in Figure~\ref{opticalspectrum}
and photometry presented in Table~\ref{lpvtable}
and Figure~\ref{lpvlightcurve}.
}
\label{opticalimages}
\end{figure}

Due to the high stellar density at the \xmm\ position of \psr\
and the possibility that the optical counterpart is highly
reddened, a definite optical/IR counterpart could not be
identified from the \xmm\ position alone (see Fig.~\ref{opticalimages}).
Two possible optical counterparts were suggested by
\citet{hal04b}, but these are $3.\!^{\prime\prime}6$
and $7.\!^{\prime\prime}7$
from the \xmm\ position, so neither was compelling.
Therefore, the astrometric accuracy of \chandra\ was brought to bear
on the problem.

\section{Chandra Observations}

Three 5~ks observations were scheduled
with the \chandra\ {\it Advanced Camera for Imaging and Spectroscopy}
\citep[ACIS-I;][]{bur97}, at intervals of 3 months,
in the expectation that at least one snapshot, or a combination of the three,
would detect enough photons from this transient source to measure its position
to sub-arcsecond accuracy and enable an optical/IR identification.
The first \chandra\ observation was performed on 2007 February 14,
and the second on 2007 May 24.
The first \chandra\ observation detected a
candidate source consisting of three photons
only $2^{\prime\prime}\!.1$ from the \xmm\
coordinates of \psr.  We reversed the pixel randomization 
that is applied by the standard data processing in order to
smooth out the gridded appearance of the image resulting from
the undersampling of the telescope PSF by the ACIS CCDs.
In this case, it is more useful to recover the most
precise positions of the three photons.  They fall within
$0^{\prime\prime}\!.7$ of each other, and have an rms dispersion in radius
of $0^{\prime\prime}\!.6$, consistent with the \chandra\ PSF.
We conclude that these three photons represent a single,
real source (see below).

The second \chandra\ observation detected 35 photons
at a position identical with the first,
R.A. = $18^{\rm h}29^{\rm m}43.\!^{\rm s}97$,
decl. = $-09^{\circ}51^{\prime}23^{\prime\prime}\!.2$ (J2000.0),
leaving no doubt that this is the counterpart of \psr. 
For each observation, the nominal 90\% position
uncertainty is $0^{\prime\prime}\!.6$, which is dominated by
the aspect solution.  Although several more
faint sources are detected in the \chandra\ images, none have obvious
optical/IR counterparts that can be used to further
refine the absolute astrometry.

Given the small number of photons detected in the first \chandra\
observation, we address here the probability that it is a real
detection, as opposed to a background fluctuation.
First, we note that the group of
three photons meets the expectation of a hard, absorbed source,
with energies of 2.3, 3.5, and 5.6~keV.  In the $2-8$~keV band,
there are a total of 29 ``events'' in a $30^{\prime\prime}$
radius around the \chandra\ position of \psr, which we take as an estimate
of the background rate for the calculation of chance coincidence.
Three of these 29 events comprise the candidate source, and fall
within the PSF of the second \chandra\ detection.
The probability that the three source
photons are a chance coincidence is
then $\approx 29\times 28\times 27\times (0.6/30)^6 = 1.4 \times 10^{-6}$.
We conclude that there is no doubt that the three-photon source is a
detection, even as its flux is uncertain by $\sim 60\%$.

The hardness ratio of the combined \chandra\ source detections is
HR=$+0.9\pm0.1$.  Here, HR=$(C_{\rm h}-C_{\rm s})/(C_{\rm h}+C_{\rm s})$,
where $C_{\rm s}$ and $C_{\rm h}$ are the counts in the
$0.5-2$ and $2-10$~keV bands, respectively.  This is consistent
with the hard, absorbed \xmm\ spectrum, and indicates that the 
neutron star is still accreting at this low level.
Assuming the \xmm\ spectral parameters, the \chandra\ count rates in the two
observations correspond to $2-10$~keV fluxes of
$2.5 \times 10^{-14}$ and
$2.8 \times 10^{-13}$ ergs~cm$^{-2}$~s$^{-1}$, respectively,
derived using the web-based simulator 
PIMMS\footnote{http://heasarc.gsfc.nasa.gov/Tools/w3pimms.html}.
Fluxes this small are to be expected from \psr,
since in two out of three \xmm\ observations,
we found an upper limit of $< 5 \times 10^{-14}$ ergs~cm$^{-2}$~s$^{-1}$.
Thus, the flux seen from \psr\ has varied by a factor of $\approx 6800$.

\begin{figure}[b]
\centerline{
\vspace{0.1in}
\includegraphics[width=3.2in]{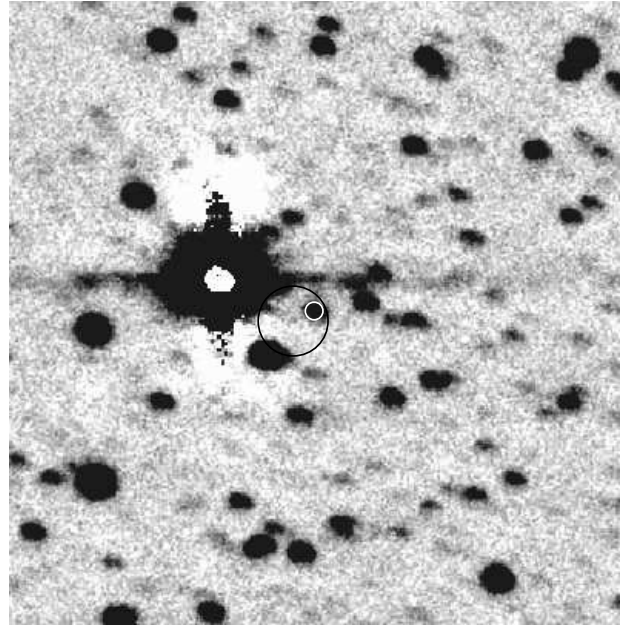}
}
\caption{$H$-band image of the field of \psr\ from the 
MDM 2.4m telescope on 2005 June 17.
North is up, and east is to the left.
The image is $0.\!^{\prime}9 \times 0.\!^{\prime}9$.
The black circle is the \xmm\ 90\% localization, of radius
$3.\!^{\prime\prime}2$, and the white circle is the
\chandra\ 90\% localization, of radius $0.\!^{\prime\prime}6$.
The probable counterpart at the \chandra\ position
has $H=13.9$ and also $K=12.7$ (the latter image not shown).
The bright, saturated star is \lpv,
a late M giant and LPV,
with optical spectrum shown in Figure~\ref{opticalspectrum}
and photometry presented in Table~\ref{lpvtable}
and Figure~\ref{lpvlightcurve}.
}
\label{irimage}
\end{figure}

\section{Optical and Infrared Observations}

We observed the location of \psr\ on a dozen epochs from 2005 April
to 2007 March using the MDM Observatory's 1.3 m McGraw-Hill telescope
\citep{weh71}.
Images were taken in the $R$ and $I$ bands using a SITe $1024\times1024$,
backside illuminated CCD with $24\mu$m ($0.\!^{\prime\prime}5$) pixels.
\citet{lan92} standard stars were used to calibrate one set of images
obtained under photometric conditions, and this calibration was
transferred to all of the other images, many of which were not photometric.
Figure~\ref{opticalimages} shows combined sets of the better images.
An astrometric solution was derived in the reference
frame of the USNO-B1.0 catalog \citep{mon03} using 29 stars that
have an rms dispersion of $0.\!^{\prime\prime}3$ about the fit,
comparable to the stated uncertainties in the catalog.  
The nearest star is $3.\!^{\prime\prime}6$ from the \xmm\
position, but inconsistent with the \chandra\ position.
Otherwise, the error circles are blank.
We derive limits of $R>23.2$ and $I>21.9$ at the \chandra\ position.

Infrared images in the $H$ and $K$ bands
were obtained on the MDM 2.4 m Hiltner telescope on 2005
June 17 using the TIFKAM imager/spectrometer \citep{pog98}
having a new $1024\times1024$ HAWAII2 HgCdTe detector array
with $0.\!^{\prime\prime}3$ pixels.
An astrometric calibration was derived
using 125 stars from the Two Micron All Sky Survey (2MASS)
with an rms dispersion of
$0^{\prime\prime}\!.14$ about the fit, comparable
to their expected uncertainties.
Figure~\ref{irimage} shows the $H$-band image,
on which a star not visible in the optical falls within
$0^{\prime\prime}\!.2$
of the \chandra\ localization of \psr.  The IR position is
R.A. = $18^{\rm h}29^{\rm m}43.\!^{\rm s}98$,
decl. = $-09^{\circ}51^{\prime}23^{\prime\prime}\!.0$ (J2000.0). 
We note that this star
is visible in 2MASS images, but not measured
in the 2MASS catalog, evidently due to the glare of the
nearby bright star.  (The special properties of the latter
are described in the Appendix).
We calibrated six stars in the field
using their 2MASS magnitudes, finding that they are internally
consistent to within $\pm 0.05$ magnitudes.  The resulting
magnitudes of the candidate counterpart are $H=13.9$ and $K=12.7$.
Therefore, we consider this a likely companion that is highly
reddened by intervening dust ($R-K>10$).

\begin{deluxetable}{lcc}
\tablecaption{Optical Photometry of \lpv }
\tablehead
{\colhead{Date} & \colhead{$R$} & \colhead{$I$}
}
\startdata
2005 Apr 23 &	$17.99 \pm 0.02$ & $15.01 \pm 0.02$ \\
2005 Aug 27 &	$16.85 \pm 0.02$ & $13.90 \pm 0.02$ \\
2005 Aug 29 &	$16.85 \pm 0.02$ & $13.89 \pm 0.02$ \\
2006 Mar 31 &	$19.48 \pm 0.07$ & $16.47 \pm 0.02$ \\
2006 Apr 02 &	$19.53 \pm 0.04$ & $16.52 \pm 0.02$ \\
2006 Apr 23 &	 ...             & $16.77 \pm 0.08$ \\
2006 Apr 29 &	 ...             & $16.80 \pm 0.08$ \\
2006 May 25 &	$19.64 \pm 0.02$ & $16.56 \pm 0.02$ \\
2006 Jun 29 &	$19.03 \pm 0.02$ & $16.10 \pm 0.02$ \\
2006 Sep 17 &	$18.15 \pm 0.02$ & $15.17 \pm 0.02$ \\
2007 Feb 23 &	$16.57 \pm 0.02$ & $13.63 \pm 0.02$ \\
2007 Mar 3 &	$16.69 \pm 0.02$ &  ...             \\
2007 Mar 8 &	$16.74 \pm 0.02$ & $13.76 \pm 0.02$ \\
\enddata
\label{lpvtable}
\end{deluxetable}

\begin{figure}[t]
\centerline{
\hfil
\includegraphics[width=2.4in,angle=-90.0]{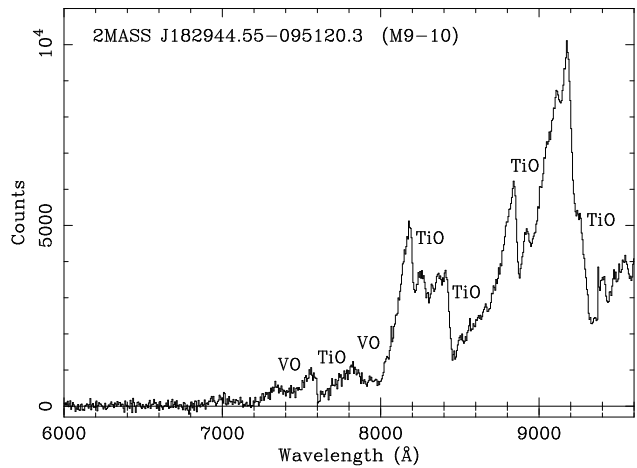}
\hfil
}
\caption{\small Optical spectrum of \lpv\
in the vicinity of \psr\
obtained on the MDM 2.4m telescope on 2005 May 6.
}
\label{opticalspectrum}
\end{figure}

\begin{figure}[b]
\vspace{0.1in}
\centerline{
\hfil
\includegraphics[width=3.2in]{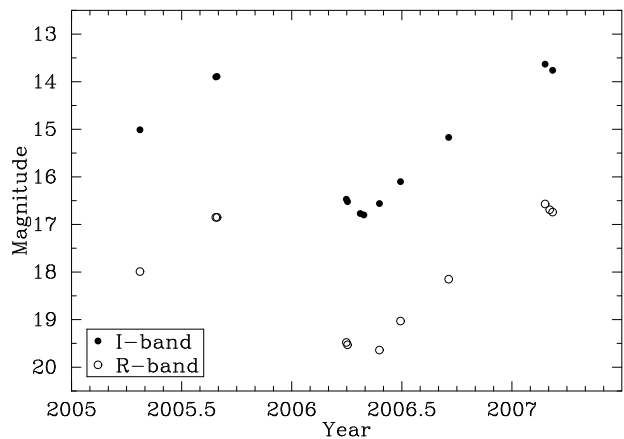}
\hfil
}
\caption{\small Photometry of \lpv\
in the vicinity of \psr\ obtained on the MDM 1.3m telescope.
Data are listed in Table~\ref{lpvtable}.
}
\label{lpvlightcurve}
\end{figure}

\section{Discussion}

Accepting the association with a bright infrared counterpart,
we rule out the hypothesis of an isolated
neutron star, specifically a transient AXP, whose periods fall
in the range 2--12~s, encompassing that of \psr.
The X-ray spectrum is also harder than that of any AXP, being similar
to high-mass binaries, and also to the LMXB pulsar
4U~1626$-$67 \citep{kii86}, as noted by \cite{mar04}.

It is unlikely that \psr\
is a Roche-lobe filling LMXB, as its ratio of IR to X-ray
luminosity is too large to be accounted for by either 
accretion-disk reprocessing or X-ray heating of the companion star's
photosphere, indicating instead that the IR emission must be
intrinsic to the companion star.  Quantitatively, if we
suppose that the IR/optical spectrum has the
form $F_{\nu} \propto \nu^{1/3}$,
typical of accretion disk emission, then the
intrinsic IR color is $(H-K)_0 = 0.44$,
requiring $A_K = 1.34$.
The intrinsic flux between the $B$ and $K$ bands is then
$\approx 1.4 \times 10^{-10}$ ergs~cm$^{-2}$~s$^{-1}$,
which is comparable to the X-ray flux in the high state.
Such optical/X-ray ratios are typical of Be X-ray binaries
or wind-fed HMXBs, as is the high-state
X-ray luminosity of $2 \times 10^{36}\ (d/10\ {\rm kpc})^2$
ergs~s$^{-1}$ (see below).

However, we cannot rule out the possibility that
the companion star is a low-mass red giant,
similar to the pulsar GX~1+4, and other recently discovered
M-giant counterparts of X-ray sources \citep{mas07}.
In either the HMXB or the M-giant case, the infrared emission
comes from the intrinsic stellar photosphere, and is too bright
to be affected by the state of the X-ray source.

The extreme amplitude of X-ray variability, by a factor
of 6800, is most typical of a Be X-ray transient,
or perhaps a wind-fed system in which the
neutron star is in an eccentric orbit.  X-ray turn-ons
may occur when the neutron star passes through the
circumstellar disk around a Be star companion or the
denser parts of a stellar wind.
Although the sparse temporal sampling of this source does
not allow a meaningful estimate of its recurrence time,
its two known outbursts were separated by $\sim 1.3$~yr,
which may be the orbital period or a multiple of it.
The relation
between orbital and spin periods of Be/X-ray binaries
\citep{cor86,lay05} would suggest an orbital period
in the range 20--60 days.
More frequent X-ray monitoring, or pulse timing in a high
state, may be able to discover the orbital period.

The IR counterpart of \psr\ appears too bright, given
its large extinction, to be a main-sequence B star.  A star
of spectral type B0V has absolute magnitude $M_K=-2.01$ \citep{han97}
and $H-K=-0.08$ \citep{win97}.  In this case, since we measure
$H=13.9$ and $K=12.7$, the required extinction $E(H-K)=1.28$
is equivalent to $A_V \approx 20$, which corresponds to
$d \sim 10$~kpc assuming $A_V/d \sim 2$ mag~kpc$^{-1}$.
At this distance, a B0V star would have apparent magnitude $K=15.2$.
However, Be stars are generally found 0.5--1 magnitudes above
the main sequence \citep{sle88}, and also up to luminosity
class III.  So we do not rule out a Be star based on
this crude argument and approximate distance.
At $d=10$~kpc, the maximum observed 2--10~keV
X-ray luminosity of \psr\ is $2 \times 10^{36}$ ergs~s$^{-1}$,
typical of Be X-ray transients and wind-fed systems,
and the minimum luminosity
is $\sim 3 \times 10^{32}$ ergs~s$^{-1}$.  At this distance, it
remains inconclusive whether absorption local to the X-ray source,
e.g., from the companion's wind, is needed to help account for the
X-ray measured column density.

Although most transient HMXBs are of the Be type,
it is possible that the companion of \psr\
is an O or B supergiant.  Many such supergiant 
X-ray pulsars having large column densities are
being found by the {\it INTEGRAL} Observatory
\citep{wal06}.  However, their spin periods are
generally longer than 100~s, while the 7.8~s period
of \psr\ is typical of a Be X-ray binary.  One possible
counterexample to this period trend is the 4.7~s
pulsar AX~J1841.0$-$0536 \citep{bam01}, optically identified
by \citet{hal04} and classified as a B0.2Ib supergiant
by \citet{neg05,neg07} \citep[see also][]{nes07}.
Further observations are needed to classify \psr.

\section {Conclusions}

We have characterized the X-ray spectrum and amplitude of
variability of the poorly studied transient X-ray pulsar
\psr, showing that it emits hard X-rays, i.e., accretes,
over a range of at least 6800 in luminosity, and has
had at least two outbursts in a period of 1.3~yr.
From the \chandra\ position, a reddened ($R-K>10$)
IR counterpart was found that is too bright
to be either an AXP or a low-mass main sequence companion.
However, we cannot determine from our
$H$ and $K$ photometry alone the spectral type of
the companion, whether Be, OB supergiant, or red giant.
The large X-ray measured $N_{\rm H}$ in excess of the
total Galactic column suggests the presence of
absorption local to the source, e.g., the wind 
or circumstellar disk of the companion star.
We estimated the distance to the source by assuming that
the companion is of spectral type B0, and translating
the $A_V=20$ implied by its $H-K$ color to $d \sim 10$~kpc.
This corresponds to a maximum X-ray luminosity of
$2 \times 10^{36}\ (d/10\ {\rm kpc})^2$ ergs~s$^{-1}$.
Most important, infrared spectroscopy is now needed to classify
the companion and obtain an actual measurement of its
distance.  

\acknowledgments

We thank Ian McGreer for obtaining the crucial infrared data at
MDM Observatory, and Pietro Reviglio for obtaining the optical
spectrum shown in Figure~\ref{opticalspectrum}.
This investigation is based on observations obtained with \xmm, an ESA
science mission with instruments and contributions directly funded by
ESA Member States and NASA.  Support for this work was provided by
the National Aeronautics and Space Administration through
{\it Chandra} Award SAO GO7-8035X
issued by the {\it Chandra} X-ray Observatory Center,
which is operated by the Smithsonian Astrophysical Observatory
for and on behalf of NASA under contract NAS8-03060.
This publication makes use of data products from the Two Micron All Sky
Survey, which is a joint project of the University of Massachusetts and
the Infrared Processing and Analysis Center/California Institute of
Technology, funded by the National Aeronautics and Space Administration
and the National Science Foundation.  We thank the anonymous referee
for useful suggestions.


\appendix

\section{A Long-Period Variable in the Vicinity of XTE J1829$-$098}

We originally suggested a very bright, red star,
\lpv, with 2MASS magnitudes $J=8.04$, $H=6.11$, and $K=4.98$
that is $7.\!^{\prime\prime}7$ from the \xmm\ position,
as a possible supergiant companion of \psr\ \citep{hal04b}.
It is the saturated star in the
infrared image of Figure~\ref{irimage}.
While we have now ruled it out, we did discover in the course
of this investigation that it is an
extreme late-type red giant or supergiant that displays
long-term photometric variability typical of such stars. 

Figure~\ref{opticalspectrum} shows a spectrum of \lpv\
obtained on the MDM 2.4m telescope on 2005 May 6.
No flux is detected from it shortward of 6800 \AA.
Longward of 6800~\AA\ are deep TiO and VO bands
indicating spectral type M9-10
\citep[see][for comparison spectra]{woo83b,flu94}.
Figure~\ref{lpvlightcurve} and Table~\ref{lpvtable}
show the photometric variability of this star in
our $R$ and $I$-band images, exceeding 3 magnitudes
with an apparent period of $\approx 1.5$~yr.  In this
study, it was important that the same detector/filter
combination was used each time, since the complex
red spectrum of \lpv\ would yield
very different photometry through slightly different bandpasses.

Red supergiant and asymptotic giant branch
(AGB) stars are long-period variables (LPVs: prototype Mira)
due to the pulsational instability of their envelopes, and they
can vary by several magnitudes on time scales of hundreds of days.
We conclude that this star is a luminous giant or supergiant
because its magnitudes indicate significant extinction,
even in the $K$ band. Its 2MASS color is $J-K=3$, while
LPVs generally have intrinsic $(J-K)_0=1.1-1.8$ \citep{woo83a}.
Adopting $(J-K)_0=1.5$, we estimate $E(B-V)\approx 3$,
$A_V \approx 9$, and thus $d\sim 4.5$~kpc assuming
$A_V/d \sim 2$ mag~kpc$^{-1}$.
In this case, the absolute $K$ magnitude is $M_K \sim -9.4$,
similar to AGB stars having
periods of $\approx 1.5$~yr \citep{woo83b}.

Until a spectral type and distance to \psr\ are determined,
we have no strong reason to suppose that \lpv\ is a
member of a stellar association that includes \psr.

\end{document}